\documentclass{svjour2}                    
\smartqed  
\usepackage{graphicx}
\usepackage{amssymb,amsfonts,amsmath}

\begin{document}

\title{Evolution of Robustness and Plasticity under Environmental Fluctuation:
Formulation in terms of Phenotypic Variances}

\author{
Kunihiko Kaneko
}
\institute{
Center for Complex Systems Biology and  Department of Basic Science, Univ. of
Tokyo, 3-8-1 Komaba, Tokyo 153-8902, Japan\\
}
\date{Received: date / Accepted: date}
\maketitle

\begin{abstract}
The characterization of plasticity, robustness, and evolvability, an important issue in biology, is studied in terms of phenotypic fluctuations. By numerically evolving gene regulatory networks, the proportionality between the phenotypic variances of epigenetic and genetic origins is confirmed. The former is given by the variance of the phenotypic fluctuation due to noise in the developmental process; and the latter, by the variance of the phenotypic fluctuation due to genetic mutation.  The relationship suggests a link between robustness to noise and to mutation, since robustness can be defined by the sharpness of the distribution of the phenotype.  Next, the proportionality between the variances is demonstrated to also hold over expressions of different genes (phenotypic traits) when the system acquires robustness through the evolution. Then, evolution under environmental variation is numerically investigated and it is found that both the adaptability to a novel environment and the robustness are made compatible when  a certain degree of phenotypic fluctuations exists due to noise. The highest adaptability is achieved at a certain noise level at which the gene expression dynamics are near the critical state to lose the robustness.
Based on our results, we revisit Waddington's canalization and genetic assimilation with regard to the two types of phenotypic fluctuations.
\keywords{Robustness, Fluctuation-Response Relationship, Evolution, Genetic Variance}

\end{abstract}

\section{Introduction}

In evolutionary biology, plasticity and robustness are considered the basic characteristics of phenotypes; these characteristics have  been widely discussed for decades.
In general, phenotypes are shaped from genotypes as a result of developmental dynamics\footnote{Here, "development" refers to a dynamic process that shapes the phenotype and is not restricted to multicellular organisms. Developmental dynamics are also observed in unicellular organisms, for instance, gene (protein) expression dynamics by regulatory networks.}, under an environmental condition.
Although these dynamics are determined by genes, they can be stochastic in nature, owing to some noise in the developmental process; further, these dynamics also depend on the environmental condition.

Plasticity refers to the changeability of the phenotype against the environmental change.
Through developmental dynamics, the influence of the environment is amplified or reduced[1-6].
In other words, plasticity concerns how the developmental dynamics are
affected by the environmental change.

Of course, the phenotype depends on the genotype.  This changeability against genetic change is called evolvability
and is related to the sensitivity of developmental dynamics against genetic change.
In this sense, both plasticity and evolvability represent the responsiveness against external perturbations and, in simple terms, a sort of "susceptibility" in statistical physics.

Robustness, on the other hand, is defined as the ability to function against possible disturbances in the system
[7-13].
Such disturbances have two distinct origins: non-genetic and genetic.
The former concerns the robustness against the stochasticity that can arise during the developmental process, while the latter
concerns the structural robustness of the phenotype, i.e., its rigidity against the genetic changes produced by mutations.
If the variance of phenotype owing to disturbances such as noise in gene expression dynamics
is smaller, then the robustness is increased.
In this sense, the variance of phenotypic fluctuations serves as a measure of robustness, as has already been discussed\cite{Plos1}.

Now, there exist certain basic questions associated with robustness and evolution[1-8].
Does robustness increase or decrease through evolution?
If it increases, the rigidity of the phenotype against perturbations also increases.
Consequently, the plasticity, as well as evolvability, may decrease with evolution. If that is the case, then the question that arises is: How can the plasticity needed to cope with a novel environment be sustained?

The decrease in plasticity and increase in robustness with evolution was actually observed
in laboratory experiments under fixed environmental and fitness conditions and was also confirmed through numerical experiments.
In a simulated evolution of catalytic reaction and gene regulatory networks under a given single fitness condition,
the fluctuation decreases through the course of evolution\cite{Plos1,JTB}.
Such networks evolve to reduce the fluctuation by noise, for a sufficient noise level.
Through this evolution, robustness against noise increases, leading to the decrease in phenotypic plasticity.
However, as a system is more and more adapted to one environment, the phenotype fitted for it would lose the potentiality to adapt to a novel environmental condition.

In the present organisms, however, neither the evolutionary potential nor the phenotypic fluctuation vanishes.
Even after evolution, the phenotype in question is not necessarily fixed at its optimal value, but its
variance often remains rather large.
There can be several sources for the deviation from the optimal state,
which is neglected in the idealized numerical and laboratory experiments with
a single fitness condition.
 One of the most typical factors for this deviation is  environmental variation.
With environmental change, the phenotype for producing the fittest state is not fixed
but may vary over generations. Then, we are faced with the following questions:
First, under environmental variation, can robustness and plasticity coexist through the course of evolution?
Second, are the phenotypic fluctuations sustained, to maintain the adaptability to environmental changes?
Third, how are the opposing features of robustness and adaptability to the new environment compromised?
Finally, is there a noise level optimal for achieving both adaptability and robustness?  We address these questions in the present paper.

\section{Background: Isogenic phenotypic fluctuation and evolution speed}

As for the fluctuation, there is an established relationship
between the evolutionary speed and the variance of the fitness, namely, the so-called fundamental theorem of natural selection proposed 
by Fisher\cite{Fisher,Futuyma,Hartl}; this theorem states that the evolution speed is proportional to $V_g$, the variance of fitness due to genetic variation.
Besides the fitness itself, any phenotypes are generally changed by the genetic variation.  Now, relationship between the evolution speed of each phenotype
with its variance due to genetic change is established as Price equation, formulated through the covariance between a phenotype and fitness\cite{Price,Lande}.
Statistical-mechanical interpretation of Fisher's theorem in terms of the fluctuation-response relationship\cite{Ao} or fluctuation theorem\cite{Mustonen}  was also discussed.

Besides the genetic change, there are sources of phenotypic fluctuations even among isogenic individuals, and under a fixed environment.
In fact, recent observations in cell biology show that there are relatively large fluctuations even among isogenic individuals. 
The protein abundances over isogenic individual cells exhibit a rather broad distribution,
i.e., the concentration of molecules exhibits quite a large variance over isogenic cells
\cite{Elowitz,Bar-Even,Kaern,Furusawa}.  This variance is due to stochastic gene expressions and to other external perturbations.
Furthermore some of such phenotypic fluctuations
are indeed tightly correlated with the fitness, and hence the fitness also shows (relatively large) isogenic fluctuations.  For example, 
large fluctuations in the growth speed (or division time) are observed among isogenic  bacterial cells\cite{Tsuru,Wakamoto}.

Hence, there are two sources of variations for the fitness or phenotype: genetic and epigenetic. The former concerns structural change in 
developmental dynamics with genetic change, and the latter concerns the noise during the gene expression dynamics.
As mentioned, the relationship between the former variance with the evolution speed was established as Fisher's theorem (for fitness)
and Price equation (for general phenotype).  Then, is there any relationship between the latter variance with the evolution speed?

At a glance, the latter variance might not seem to be relevant to evolution, since this epigenetic change itself due to noise is not inherited to the
offspring, in general. This is not the case.  Here, it should be noted that the degree of variance itself is a nature of developmental dynamics to shape the phenotype,
and thus depends on genotype, and can be inherited.   Hence, there may exist some correlation between evolution speed and
this isogenic phenotypic variance, which is denoted as $V_{ip}$ here
\footnote{This is not standard terminology. Phenotypic variance by non-genetic origins is often termed as environmental variation $V_e$. Here, however, the source of the variance is not necessarily environmental change but, primarily, noise in the developmental process; hence, we use a different term.  Note that the total phenotypic variance is $V_{ip}+V_g$ under a certain ideal condition\cite{Futuyma,Hartl}, if they are added independently.}.

Indeed, from evolution experiments for increasing the fluorescence in an inserted protein in bacteria\cite{Sato}
and also from numerical experiments for evolving reaction networks or gene regulatory networks\cite{JTB,Plos1}
to increase a given fitness, we have observed that

$V_{ip} \propto $ Evolution Speed,

\noindent
where the evolution speed is defined as the increase of the fitness.
In the experiment, for each mutant bacteria,
average fluorescence of isogenic cells was measured and that with highest average fluorescence was selected.  Simultaneously, 
the variance of isogenic bacterial population was measured to obtain $V_{ip}$.   The same procedure was applied to measure the evolution speed
and the fitness variance in numerical evolution to increase a given fitness.  Interestingly,
these experiments support the above relationship, at least approximately.  Some 'phenomenological explanation' 
was proposed by assuming a  Gaussian-type distribution $P(x;a)$ of the fitness (phenotype) $x$ as parameterized by a "genetic" parameter, $a$.  
and a linear change in the peak position of $x$ against $a$\cite{book,Sato,LehnerKK}. It should be noted, however, that the argument based on 
the distribution $P(x;a)$ is not a "derivation" but rather a phenomenological description.  Indeed, the description by $P(x;a)$ itself is an assumption: for example,
whether a genotype is represented by a scalar parameter $a$ is an assumption.

%
%

Now, considering the established Fisher's relationship, the above relationship suggests the proportionality between
$V_{ip}$ and $V_g$ through an evolutionary course.  Indeed, this relationship was  confirmed from
several simulations of models.  Again, this relationship is not derived from established relationships in population genetics.  Indeed,
the proportionality between the two is not observed in the first few generations, but observed after robust evolution preserving a single
peak in the fitness is progressed.  Considering this observation,
we previously discussed the relationship, by postulating 
evolutionary stability of the distribution over phenotype and genotype\cite{Plos1,JTB,Kaneko2009}.

The relationship between $V_{ip}$ and $V_g$  also suggests a possible link between developmental robustness to noise and evolutionary robustness to genetic changes
(mutation).  For this link, we first note that the two types of variances $V_{g}$ and $V_{ip}$ lead to two kinds of robustness:  
rigidity of the fitness (phenotype) against genetic changes produced by mutations and robustness against the stochasticity 
that can arise in an environment or during the developmental process.
When the fitness (phenotype) is robust to noise in developmental process, it is rather insensitive to the noise,
and therefore, its distribution is sharper. Hence, the (inverse of the) variance of isogenic distribution,
$V_{ip}$, gives an index for robustness to noise in developmental dynamics\cite{Plos1,JTB}.
On the other hand, if $V_g$ is smaller, the phenotype is rather insensitive to genetic changes, implying higher genetic (or mutational) robustness.
Hence the correlation between $V_{ip}$ and $V_g$ implies correlation between the two types of robustness.  Indeed,
previous simulations suggest that robustness to noise fosters robustness to mutation.  Congruence between evolutionary and developmental
robustness was also discussed by Ancel and Fontana for the evolution of RNA\cite{AncelFontana}.

To close this section, a brief remark on the measurement of $V_g$ is given here.
Because the fitness (phenotype) distribution exists even in isogenic individuals, the variance of its distribution over a heterogenic 
population includes both the variance among isogenic individuals and that due to genetic variation. 
To distinguish between the two contributions, we first measure the average fitness (phenotype) over isogenic individuals and then compute 
the variance of this average over a heterogenic population. This variance will only be attributed to genetic heterogeneity.
This variance is denoted by $V_g$\footnote{According to the conventional terminology in population genetics,
this variance is referred to as "additive" genetic variance;
The term "additive" is included, to remove the variance due to sexual recombination.  Here we
do not discuss the influence of recombination and, thus, do not need to distinguish genetic variance from  the additive one.}. 

\section{Our Standpoint and Model}

\subsection{Dynamical-systems approach to evolution-development relationship}
  
To discuss the evolution, the fitness landscape as a function of genotype is often adopted following Wright's picture\cite{Wright}.  
Energy-like fitness function is assigned to a genotype space, i.e., $Fitness=f (Genotype)$.  Here the evolution is discussed as a hill-climbing process 
through random change in genotype space and selection.  Although this viewpoint has been important in evolutionary studies,
another facet in evolution has to be also considered, to discuss the phenotypic plasticity and evolution-development relationship, that is
genotype-phenotype mapping shaped by the developmental dynamics.  Here we focus on this evolution-development relationship.

Note that the fitness is a function of (some) phenotypes (say a set of protein abundances).  The phenotypes are determined by developmental
dynamics (say gene expression dynamics), whose rule (e.g., set of equations or parameters therein) is
governed by genes. Thus this evolution-development scheme is represented as follows:

(i) {\bf Fitness= F(phenotype)}

(ii) {\bf Phenotype} determined as a result of {\bf developmental dynamics}

(iii) {\bf Rule} of {\bf developmental dynamics} given by {\bf genotype}

According to (ii) and (iii), genotype-phenotype mapping is shaped, which is not necessarily deterministic.  As discussed,
the phenotypes from isogenic distribution are distributed, since the dynamics (ii) involve stochasticity (in gene expression).  
Indeed, with this stochasticty,
reached attractors by the dynamics are not necessarily identical, and accordingly the fitness may vary even among the isogenic individuals,
which lead to the variance $V_{ip}$.  If the dynamics to shape the phenotype is very stable, the variance $V_{ip}$ is smaller.
Further, the degree of the change in phenotype by the genetic change depends both on (ii) and (iii), and
$V_g$ depends on the sensitivity of the phenotype to the change in the rule of the dynamics.
With the evolutionary process the genotype, i.e., the rule of the dynamics, changes,
so that $V_{ip}$ and $V_g$ change with the evolution.

The developmental dynamics, in general, involve a large number of variables (e.g., proteins expressed by genes), and are complex.  Accordingly, genotype-phenotype mapping is generally complex, and
the mapping from genotype to the fitness is not simple, even if the function in  (i) is simple.  
In several studies with adaptive fitness landscape, complexity in the fitness landscape has been taken into account, while the developmental
dynamics (ii) are not.  We take a simple fitness function (say, the number of expressed genes), but
instead take complex developmental dynamics into account, to discuss the phenotypic plasticity in developmental
and evolutionary dynamics. 

If we are concerned only with the fitness landscape (i), we can introduce a single 'energy'-like function for it
and formulate the adaptive evolution in terms of standard statistical mechanics.  Here, however,
we need to consider the dynamics to shape the phenotype (ii) \cite{AncelFontana,Ciliberti,Plos1,JTB}. 
If we adopt statistical-mechanical formulation, we need two 'energy'-like functions,
one for fitness and the other for Hamiltonian for the development dynamics, as is formulated by two-temperature statistical physics\cite{Sakata,Sakata2}.

So far, such study on evolution-development (so called 'evo-devo')
lacks mathematically sophisticated formulation as compared with the celebrated population genetics, and thus
we sometimes have to resort to heuristic studies based on numerical evolution experiments, from which
we extract 'generic' properties.
We then provide some plausible arguments (or phenomenological theory), while
 mathematical or statistical-physical formulation has to be pursued in future.

\subsection{Gene expression network model}

Following the argument in the lase section and
to discuss the issue of plasticity and robustness in genotype-phenotype mapping,  we adopted a simple model\cite{Plos1,Chaos,BMC}  for gene expression dynamics with a sigmoid input-output
behavior\cite{gene-net,Mjolsness,Sole}; it should be noted, however, that several other
simulations in the form of other biological networks
will give essentially the same result. In this model, the dynamics of a given gene expression level, $x_i$, is described by the following:

\begin{equation}
dx_i/dt =\gamma\{ f( \sum_{j}^{M} J_{ij}x_j)-x_i  \}+\sigma \eta _i(t),
\end{equation}
\noindent
where $J_{ij}=-1,1,0$. The noise term $\eta_i(t)$ is due to stochasticity in gene expression. For simplicity, it is taken to be Gaussian white noise given by
$<\eta _i(t)\eta _j(t')>= \delta _{i,j}\delta(t-t')$, while the (qualitative) results to be discussed below is
independent of the choice \footnote{Indeed, multiplicative noise depending on $x_i$, as well as a stochastic reaction
model simulated with the use of Gillespie algorithm gives a qualitatively same behavior\cite{Luis}. 
Of course the magnitude of the phenotype variance as well as the threshold noise level for error catastrophe, to be discussed below, 
depends on the form of noise. However, relationship of such threshold noise level with $V_{g}/V_{ip}$, as well as the proportionality between $V_{ip}$ and $V_g$ does not depend on the specific form of noise.}.
The amplitude of noise strength is given by $\sigma$,
which determines stochasticity in gene expression.
$M$ denotes the total number of genes; and $k$, the number of target genes that determine fitness.
Here, the function $f(x)$ represents a threshold function for gene expression.
Previously, we adopted the model (1) with

\begin{equation}
f(x)=\tanh(\beta x),
\end{equation}
where the gene expression is "on" if $x>0$ and "off" if $x<0$.

To discuss the environmental condition, however, it is relevant to introduce an "input" term
for expressing some genes. To make this input effective, we modify the function (3) as

\begin{equation}
f(x)=1/(exp(-\beta( x-\theta _i)+1) +\delta
\end{equation}

Here, $x$ is positive and is scaled so that the maximal level expression is $\sim 1$; $\delta$ takes a small positive
value corresponding to a spontaneous expression level.  If the input sum from other genes
$\sum_{j}^{M} J_{ij}x_j$ to gene $i$ exceeds the threshold $\theta_i$, the gene (protein) $i$ is expressed so that $x_i \sim 1$, as $f$ approaches a step function for sufficiently large $\beta(>1)$, which corresponds to the Hill coefficient in cell biology.
Now, if all $x_i$'s are initially smaller than the threshold $\theta_i$, they remains so if $\delta < \theta_i$. Hence, to generate some expression patterns, some input term is needed.  Here, we introduce "input" genes $j=1,...k_{inp}$, in which $x_j$ is set at

\begin{equation}
x_j = I_j  >0 \hspace{.1cm} (j=1,2,...,k_{inp}),
\end{equation}

\noindent
where $I_j$ is an external input to a set of "input genes."  These inputs are needed to express genes; otherwise, the expression levels remain sub-threshold.
In this case with eq. (3), the gene is expressed if $x_i>\theta_i$,
The initial condition is given by a state where none of the genes are expressed, i.e., $x_i\sim 0$.

Next, we set the fitness for evolution.  The fitness, $F$, is determined by whether the expressions of the given 
"target" genes are expressed after a sufficient time.
This fitness condition is given such that $k$ target genes are "on" (expressed),
i.e.,  $x_i>\theta_i$ for $M-k<i \leq M$.
Because the model includes a noise component, the fitness can fluctuate at each run,
which leads to a distribution in $F$ and $x_i$,
even among individuals sharing the same gene regulatory network.
For each network, we compute the average fitness $\overline{F}$ over $L$ runs as well as the variance.  This variance over $L$ identical individuals (having the identical networks $J_{ij}$) due to noise is $V_{ip}$, as it represents
the fluctuation of the fitness over isogenic individuals (isogenic phenotypic variance). $V_g$, to be discussed below, is the variance of $\overline{F}$ over $N$ heterogenic individuals (having different networks $J_{ij}$).

Here $V_{ip}$ is not the variance over time in a single run, but the variance over runs for identical individuals. 
With developmental process by gene expression dynamics from a given initial condition, the gene expression pattern ($x(j);j=1,..,M$)
reaches an attractor.
Reached attractors can be different by each run, due to the stochasticity in expression dynamics during the transient time steps to reach the attractor.
The fitness is computed on this attractor, which is also distributed as a result of noise.
Once the attractor is reached, the fluctuation of the fitness around it is negligible, since the fitness is not given directly by $x_i$ but defined through
a threshold function of $x_i-\theta_i$.  The variance is smaller as more runs result in the same attractor (or attractors of the same fitness)
under noise. If gene expression dynamics with such global attraction to the same-fitness attractor are shaped through the evolution,
the variance $V_{ip}$ is smaller.

Now, at each generation, there are $N$ individuals with slightly different gene regulatory networks $\{ J_{ij} \}$, and their average fitness $\overline{F}$ may differ by each.
Among the networks, we select the ones with higher fitness values.
From each of the selected networks, $J_{ij}$ is "mutated," i.e.,
$J_{ij}$ for a certain pair $i,j$ selected randomly with a certain fraction switches to one of the values among $\pm1,0$.
$N_s (<N)$ networks with higher $\overline{F}$ values are selected,
each of which produces $N/N_s$ mutants. We repeat this selection-mutation process over generations.
(For example, we choose $N = L = 500$ or 700 for most simulations, and $N_s = N/4$; the conclusion, to be shown below, does not change as long as
these values are sufficiently large. We use $\beta = 7$, $\gamma =.1$, $M = 64$, and $k = 8$
and initially choose $J_{ij}$ randomly. The probability is taken to be 1/4 for $J_{ij}=\pm 1$
and 1/2 for 0 for most simulations below, but
the results below are independent of this choice\cite{BMC}.

Now, we have two types of variances.  Besides $V_{ip}$, $V_g$ is defined as  the variance of $\overline{F}$ over the $N$ individuals having different genes (gene regulatory networks).  This shows the variance due to genetic change. 
As $V_g$ is decreased, the fitness becomes insensitive to the genetic change, i.e., the robustness to mutation is increased.  On the other hand, as $V_{ip}$ is decreased, the robustness to noise is increased, since $V_{ip}$ measures the variance due to noise in a developmental process. 
(It should be remarked that the absolute value of the fitness $F$ is not important, since the model behavior is invariant against the transformation 
$F \rightarrow c\times F$. Accordingly the absolute values of $V_{ip}$ and $V_g$ are not important, but the ratio between the two or relative change 
of the variances over generations is important. 
On the other hand, each expression $x(i)$ is already scaled so that the maximal expression is $\sim 1$).

\section{Confirmation of Relationships in Phenotypic Fluctuations}

\subsection{Proportionality between the phenotypic variances by noise and by mutation}

\begin{figure}[tbp]
\begin{center}
\includegraphics[width=9.0cm,height=7.0cm]{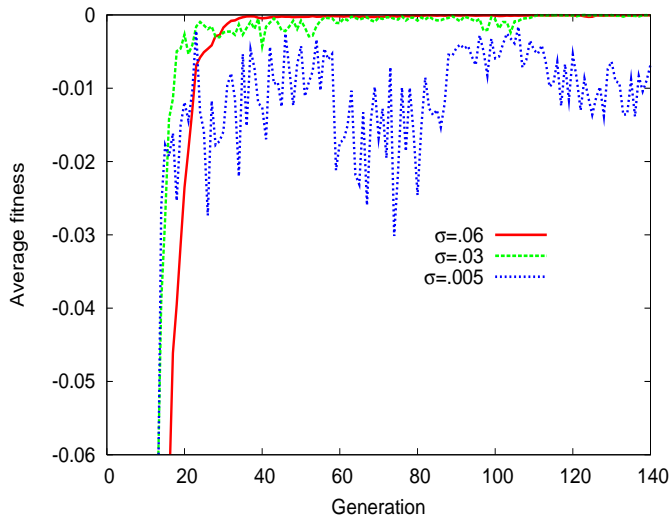}
\caption{
Evolutionary time course of the
average fitness $<\overline{F}>$.
The average of mean fitness $<\overline{F}>$ over all $N=500$
individuals that have different genotypes (i.e., networks $J_{ij}$)
in each generation is plotted.  The mean fitness of each genotype
is computed from $L=500$ runs. The plotted points are for different
values of noise strength, $\sigma=$0.005, 0.03, 0.06,
with different colors. For Figs.1-3, we choose
$M=64$, $k=8$, and $k_{inp}=8$ with $I_j=1$, while $\theta_i$ is distributed
in [0.1,.0.3].
}
\end{center}
\end{figure}

\begin{figure}[tbp]
\begin{center}
\includegraphics[width=8.5cm,height=6.5cm]{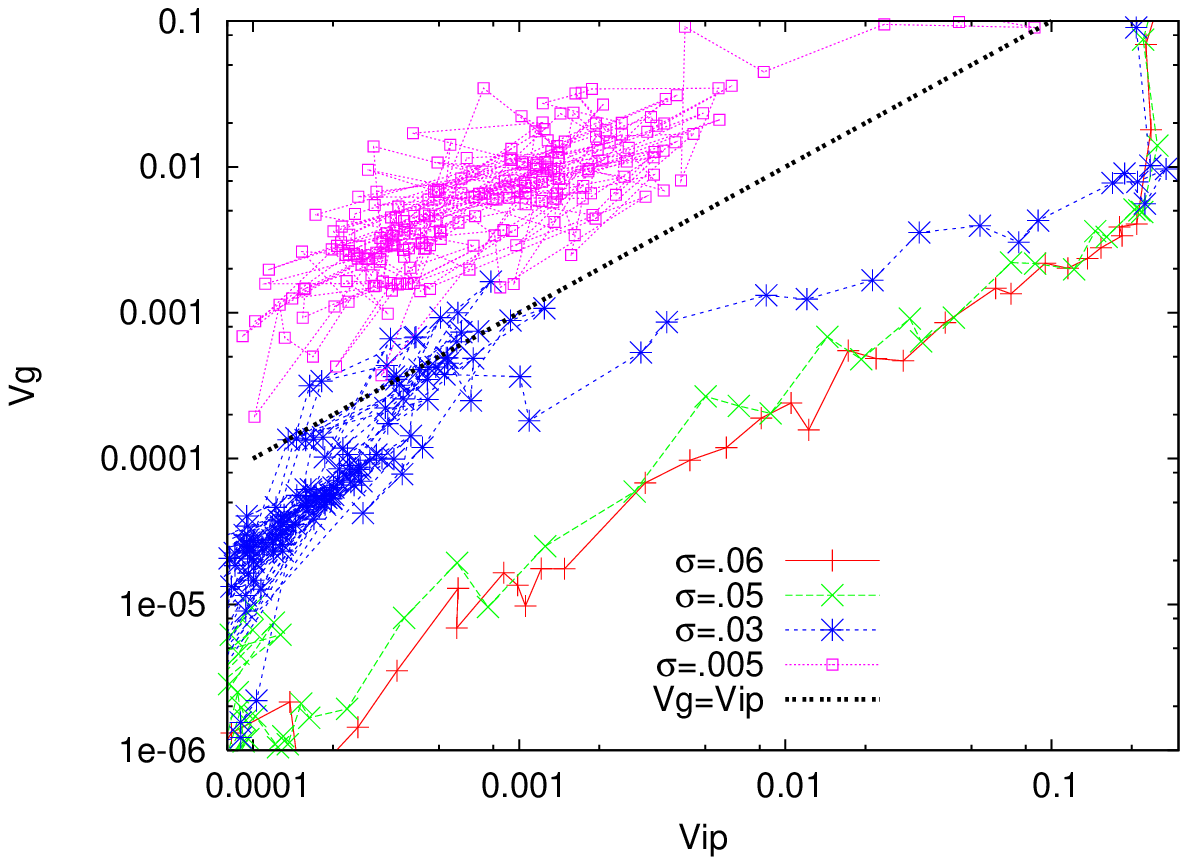}
\caption{
Relationship between $V_g$ and $V_{ip}$, the variances of the fitness.
$V_g$ is computed from $P(\overline{F})$, the distribution of mean fitness $\overline{F}$
and $V_{ip}$ is computed from the isogenic variance of the fitness over $L=500$ runs, for each genotype, and then
these values are averaged over all existing individuals. (We also
confirmed the overall relationship by using the variance for a gene regulatory network that gives the
peak fitness value in $P(\overline{F})$). The plotted points are over 140 generations.  
$\sigma=0.005$ ($\Box$), 0.03 ($*$), 0.05 ($\times$) and 0.06 ($+$).
For $\sigma > \sigma_c \approx 0.02$, both the variances decrease with generatons, so that
the right-top is the first generation and left-bottom is the 140th generation. 
After initial few generations both decrease roughly in proportion, while
at $\sigma =0.03$, the deviation from the proportionality is larger possibly because the noise level is near the critical point
to lose the robust evolutionary process.  For $\sigma=0.005$, the decrease stops after 20 generations, and the
variance values scatter for later generations.
}
\end{center}
\end{figure}

From the numerical simulation of the model, we have confirmed the following results.

(1) There is a certain threshold noise level $\sigma_c$ beyond which the evolution of robustness progresses, so that both $V_g$ and $V_{ip}$ decrease. Here, most of the individuals take the highest fitness value. In contrast, for a lower noise level $\sigma < \sigma_c$, mutants that have very low fitness values always remain. Several individuals take the highest fitness value, whereas the fraction of individuals with much lower fitness values does not decrease; hence, $V_g$ remains large (see Figs.1 and 2).

(2) At around the threshold noise level,
$V_{g}$ approaches $V_{ip}$. For $\sigma<\sigma_c$,  $V_{g}\sim V_{ip}$ holds, whereas
for $\sigma>\sigma_c$, $V_{ip}>V_{g}$ is satisfied. For robust evolution to progress, this inequality is satisfied.

(3) When the noise is larger than this threshold, the two variances decrease, while $V_{g} \propto V_{ip}$
 is maintained through the evolution course.
Hence, the proportionality between the two variances is confirmed.

Why does the system not maintain the highest fitness state under a small
phenotypic noise level with $\sigma<\sigma_c$?  Indeed, the dynamics of the top-fitness networks that evolved
under such low noise levels have distinguishable features from those that evolved under high noise levels. It was found that
for networks evolved under $\sigma>\sigma_c$, a large portion of the initial
conditions reached attractors that give the highest fitness values, whereas for networks evolved under
$\sigma<\sigma_c$, only a tiny fraction (i.e., in the vicinity of the all-off states) reached such attractors.

In other words, for $\sigma>\sigma_c$, the "developmental" dynamics that give a functional phenotype have
a global, smooth attraction to the target.
In fact, such types of developmental dynamics with global attraction are known to be ubiquitous in protein folding dynamics \cite{Onuchic,Abe},
gene expression dynamics\cite{Li}, and so forth.
On the other hand, the landscape evolved at $\sigma<\sigma_c$ is rugged. Except for the vicinity of the given initial
conditions, the expression dynamics do not reach the target pattern.

The observed proportionality between $V_{ip}$ and $V_g$ is not self-evident. Indeed, if a random network is considered for a gene regulatory network, such proportionality is not
observed. In the present simulation, after a few generations of evolution, both the variances decrease,
following proportionality, if $\sigma >\sigma_c$.
Although there is no complete derivation for this relationship,
it is suggested that this proportionality as well as the relationship $V_{ip}>V_g$ is a consequence of evolutionary stability to keep a single-peakedness in the distribution
$P(x=phenotype, a=genotype)$, under conditions of strong selective pressure and low mutation rate\cite{JTB,Kaneko2009}.

\subsection{Proportionality between the two variances across genes}

As mentioned above, the gene expression dynamics evolved under a sufficient level of noise have a characteristic property; the attractor providing the phenotype of the
highest fitness has a large basin volume and is, hence, attracted globally by a developmental process under noise.  

Note that the expression level $x_j$ of non-target genes $j$ could be either on or off, because
there is no selection pressure directed at fixing their expression level.
Still, each expression level $x_j$ can have some correlation with the fitness in general.  Hence it is also interesting to
study the variance of each expression level and discuss its evolutionary changes.

Similar to the variances for the fitness, the phenotypic variance $V_{ip}(i)$ for each gene $i$ in an isogenic population is defined on the basis of the
variance of the expression of each gene $i$, with each $X_i=Sign(x_i -\theta_i)$, in an isogenic population.
Accordingly, the variance computed by using the distribution of $\overline{X_i}$
in this heterogenic population gives  $V_g(i)$ for each gene $i$\footnote{Throughout the present paper, $V_g(i)$ and $V_{ip}(i)$ with $(i)$ denote
the variances of each phenotype, expression level $i$, while, $V_g$ and $V_{ip}$ denote the variances of the fitness.}.

\begin{figure}[tbp]
\begin{center}
\includegraphics[width=9.0cm,height=6.5cm]{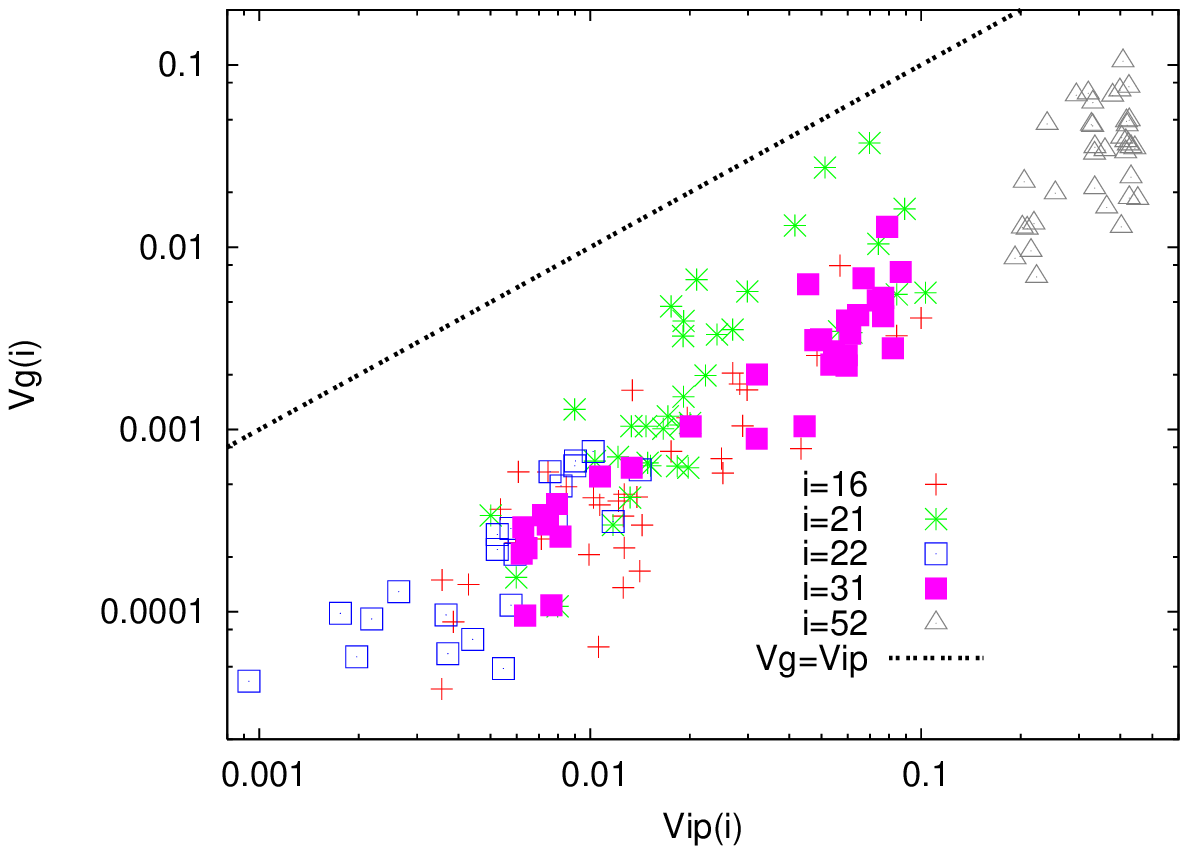}
\includegraphics[width=9.0cm,height=6.5cm]{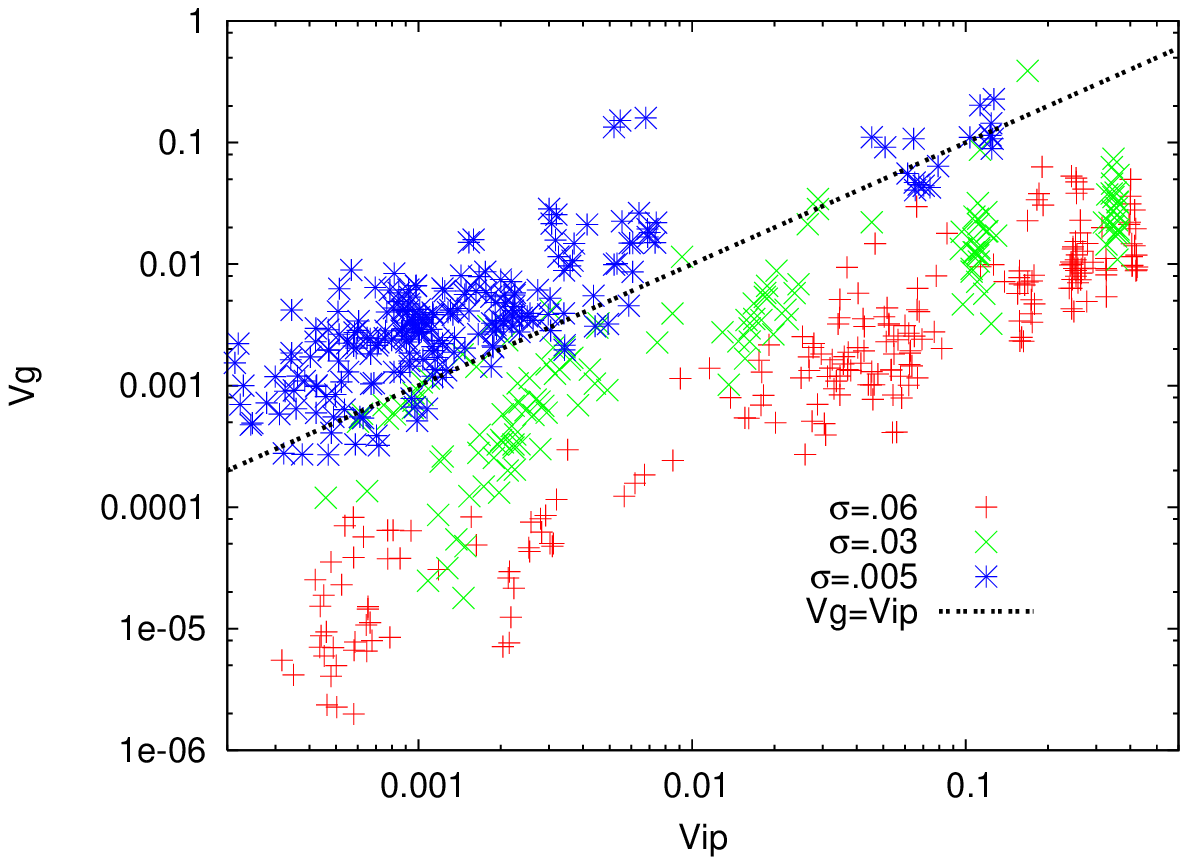}
\caption{(a) Plot of $(V_{ip}(i), V_g(i))$ for the genes $i=16,21,22,31,52$ for the generations 5-40.
Each of the variances decreases over generations, so that variance for each gene changes from right (upper) to left (lower) in the
figure.  As described in the text,
$V_{ip}(i)$ was computed as the variance of the distribution of $Sign(x_i)$
over $L=500$ runs for an identical genotype,
while $V_g(i)$ was computed as a variance of the distribution of
$(\overline{Sign(x_i)})$ over $N=500$ individuals, where
$\overline{Sign(x_i)}$ refers to the mean over 500 runs.
With generations, both the variances decrease roughly with proportion, with a trend of common proportion
coefficient.
(b) Plot of $(V_{ip}(i), V_g(i))$ across all expressed genes $i$ (i.e. for such genes $i$ that $\overline{x(i)}> \theta _i$),  
after evolution is completed (for the generations 25-60), for three values of noise levels
$\sigma=0.005$ (*), 0.03 ($\times$), and 0.06 ($+$).
}
\end{center}
\end{figure}

Following Price equation\cite{Price},
the rate of change in each expression level between generations is expected to be correlated with $V_g(i)$,
as it has direct or indirect influence to the fitness, through the gene expression dynamics (1).
In contrast, here, we are interested in the variance of isogenic phenotypic fluctuations of each expression level $V_{ip}(i)$.  
Indeed, this variance decreases over generations for most genes. As the evolution progresses, these expressions also start to be rigidly fixed so that their 
variances decrease over most genes.  In Fig.3(a), we have plotted $V_{g}(i)$ versus $V_{ip}(i)$ over generations for several genes $i$.  
We can see that they decrease (roughly) in proportion, over generations.  Furthermore, the proportion coefficient $V_g(i)/V_{ip}(i)$ 
seems to take close values across many genes.

In Fig.3(b)  we have plotted ($V_{ip}(i)$, $V_{g}(i)$), across all expressed genes, after evolution reached the genotype with the highest fitness.
As shown, the proportionality (or strong correlation) between $V_{g}(i)$ and $V_{ip}(i)$  holds
 across many (expressed) genes for a system through evolution.
The ratio $\rho=V_g(i)/V_{ip}(i)$ increases with the decrease in $\sigma$, and at around $\sigma_c$, it approaches $\sim 1$.

Although the origin of the proportionality has not yet been completely understood,
a heuristic argument is proposed by using the distribution $P(x_i,a)$ for each gene expression $x_i$ and by further
assuming that the distribution maintains a single peak up to a common mutation rate (i.e., a common mutation rate for the error catastrophe threshold) over
genes\cite{BMC}.  The latter hypothesis may be justified as a highly robust system: In such a system
with  increased error-threshold mutation rates, once the error occurs, it propagates  and percolates to many genes, so that a common
error threshold value is expected.  Note that there are some preliminary experimental supports on this proportionality of the two variances
over genes or phenotypic traits\cite{LehnerKK,Laundry,LehnerPLoS,Stearns}, although future studies are required for the confirmation.

\section{Phenotypic Evolution under Environmental Variation}

\subsection{Restoration of plasticity with the increase in fluctuations by environmental change}

\begin{figure}[tbp]
\begin{center}
\includegraphics[width=8.4cm,height=6.5cm]{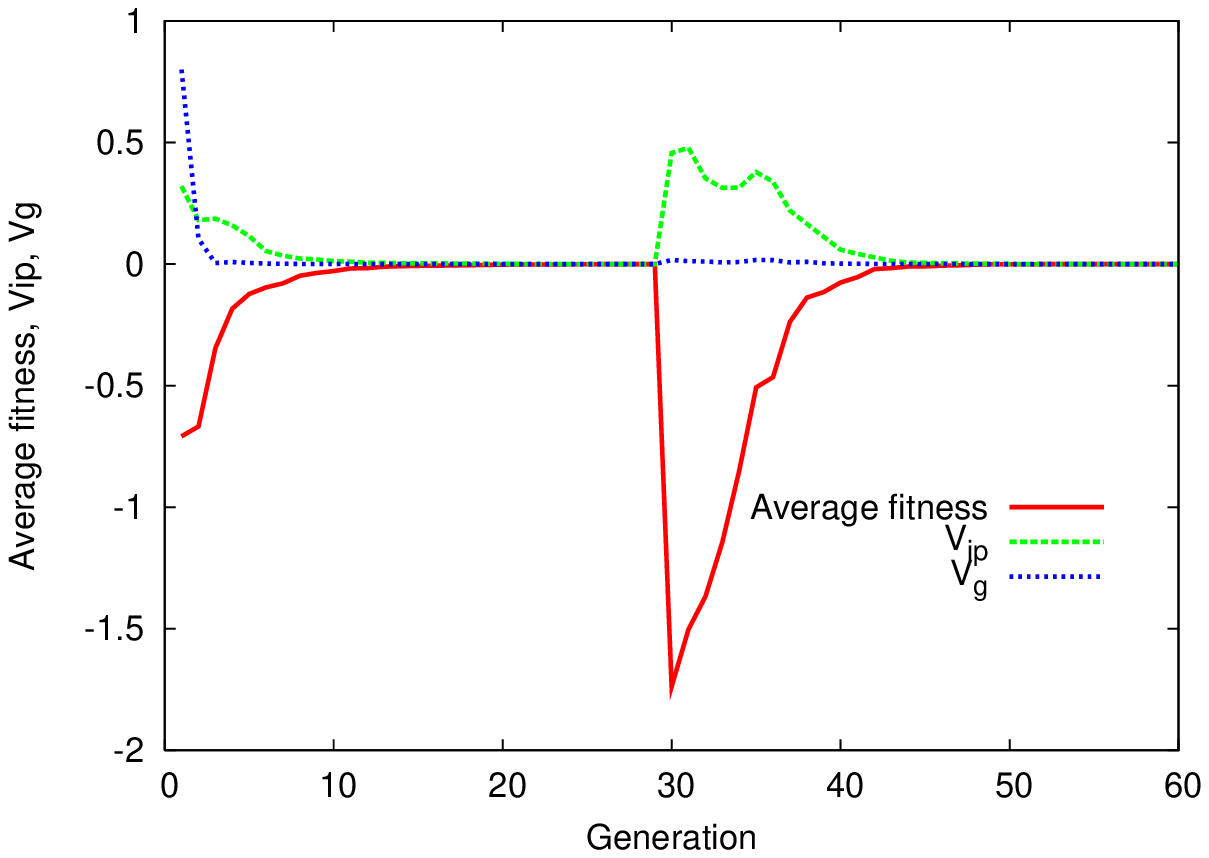}
\includegraphics[width=9.2cm,height=7cm]{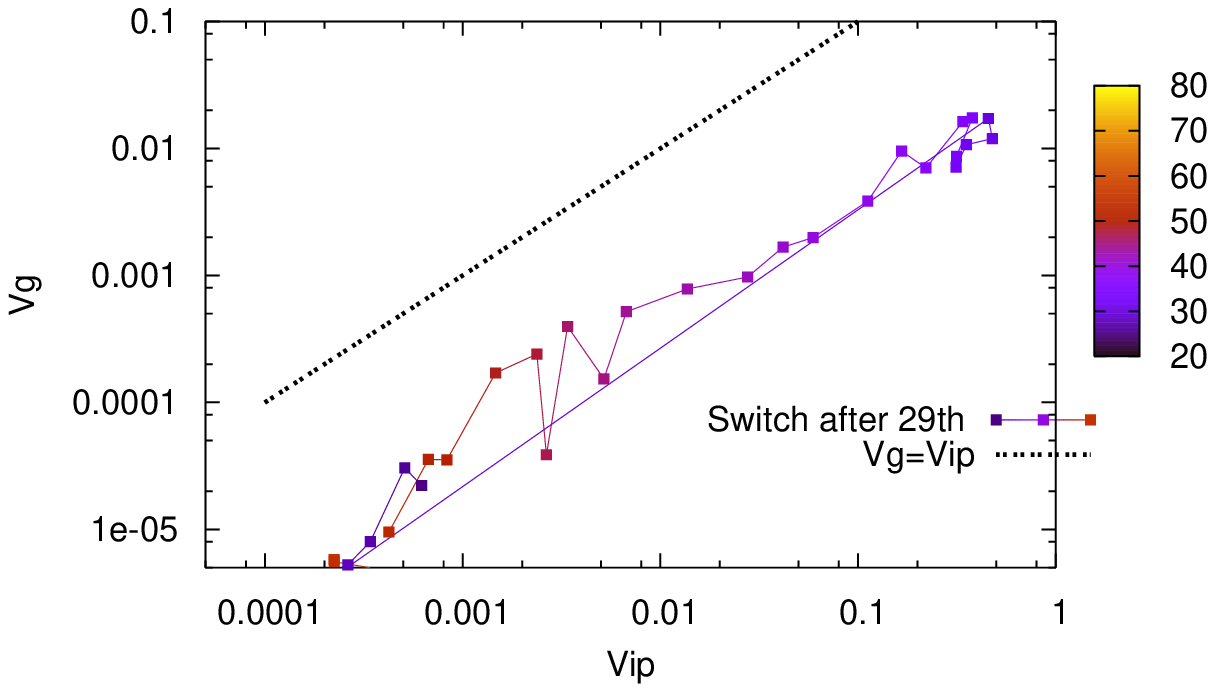}
\caption{
The time course in the fitness and the variance of the fitness over generations.
First, the evolution under an environment
$I_j=1$ for $1 \leq j \leq 4$, and $I_j=0$ for $5 \leq j \leq k_{inp}=8$ is simulated to progress up to 30 generations.
After the 29th generation, we switch the fitness condition to
$I_j=0$ for $1 \leq j \leq 4$, and $I_j=1$ for $5 \leq j \leq 8$, which is maintained at later generations.
$M=64$, $L=N=700$, and $\theta_i$ is distributed uniformly in [0.1, 0.3].
The switch initially causes a decrease in the fitness, but after a few dozens of generations, almost
all networks evolve to adapt to the new fitness condition. The noise level is set at  $\sigma =0.06 > \sigma_c$.
Top: The time course of the average fitness and the variance $V_{ip}$ throughout the evolution.
Bottom: The plot of the variances of the fitness, $V_g$ versus $V_{ip}$ for each generation after the switch of the environmental condition.
The generations (up to 60) are represented by different colors. Both the variances increase in correlation, after the switch, and later, they decrease in proportion,
to adapt to the new condition.
}
\end{center}
\end{figure}

\begin{figure}[tbp]
\begin{center}
\includegraphics[width=9.0cm,height=7.0cm]{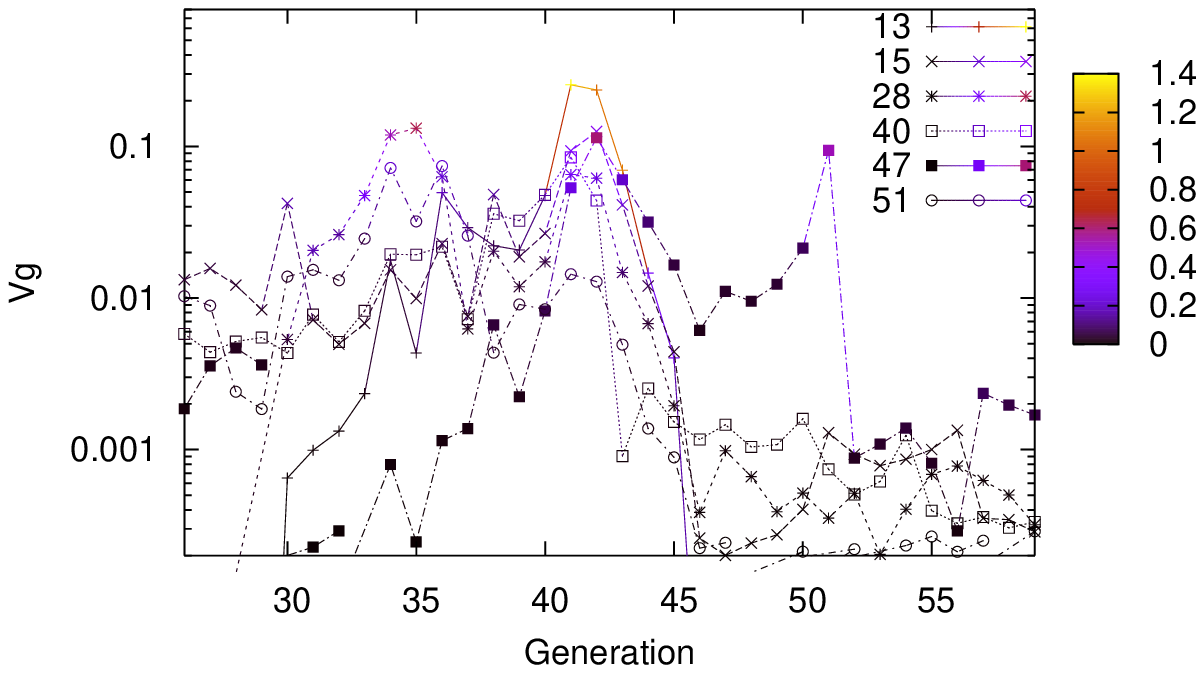}
\caption{Change in the variances $V_{ip}(i)$ and $V_g(i)/V_{ip}(i)$ at after the switch in environmental condition after the 29th generation,
as given in Fig. 4.
The variance $V_{ip}(i))$ is plotted up to generation 60, where the switch is given at the 30th generation.  Plotted for the genes $i=$13,15,28,40,47,51. 
For the genes 13, 28, and 47, $V_g(i)$ before the switch after 29th generation is smaller than $10^{-4}$. 
The color represents $V_g(i)/V_{ip}(i)$, as shown in the right bar.
}
\end{center}
\end{figure}

So far, through the selection process under a fixed fitness$/$environmental condition, both the fluctuations
and the rate of evolution decrease.  The system loses plasticity against the change caused by external noise or external
mutation.  Nevertheless, in nature, neither
the fluctuations nor the evolution potential vanish.  How are phenotypic plasticity,
fluctuations, and evolutionary potential sustained in nature?

One possible origin for the preservation of plasticity may be environmental fluctuation\cite{envfluct1}, as
has also been studied in terms of statistical physics\cite{Kussel,envfluct2,Mustonen2}.
The plasticity of a biological system is relevant for coping with the environmental change that may
alter phenotypic dynamics in order to achieve a higher fitness.
In the present modeling, there can be two ways to include such an environmental change.

One is a direct method, in which an input term given in eq.(4) is changed with each generation, while the fitness condition is maintained.
The other method is indirect, in which environmental change is introduced as the change in the fitness condition, while preserving the dynamics itself.
Here, we discuss the simulation result of the former procedure first and will consider the result from the other procedure later in \S 5.3.

To change the environmental condition, we varied the input pattern at some generation.
Here, we change the input pattern $I_j$ ($1\leq j \leq k_{inp}$) from the generation at which the system had already adapted to the environment
and decreased the phenotypic variances.
An example is plotted in Fig.4. Here, $I_j$ initially takes
$I_j=1$ for $1\leq  j \leq  k_{inp}/2$ and $I_j=0$ for $k_{inp}/2<j \leq k_{inp}$, before switching
to $1-I_j$ after the 29th generation.
By switching the environment, the fitness first decreases and later adapts to the new environment (Fig.4a).

To determine the evolution of phenotypic plasticity, we computed the variances of the fitness, $V_{ip}$ and $V_g$,
over successive generations (see Fig.4b).
After the switch of the fitness condition, both $V_{ip}$ and $V_g$ first increase to a relatively high
level and continuously increasing further over a few generations.
At later generations, both $V_{ip}$ and $V_g$ again decrease, maintaining the proportionality.
The proportionality law between the genetic and epigenetic variances
is satisfied with both increase and decrease in plasticity through the evolution.

With the increase in $V_{ip}$, the fitness is more variable with noise, which also leads to higher changeability against environmental conditions.
The gene expression dynamics regain plasticity, which allows for the switch of the target genes after further generations.
Then, with the increase in $V_g$, the changeability against genetic change increases, thus increasing the evolvability.

Next, we explore the change in the variances of the expression of each gene. As shown in Fig.5,
both variances $V_{ip}(i)$ and $V_g(i)$ increase, as a result of environmental change.
Expressions of all genes are more variable against noise and mutation.
Most gene expressions gain higher plasticity by increasing the variance in their expression.
Besides the increase in the variances, the ratio 
$\rho_i = V_{g}(i)/V_{ip}(i)$ also increase at some generation, to approach  $V_{ip}\sim V_g$ (see the color change
in Fig.5, where the red color shows higher ratio $\rho_i$. 
This increase in $V_g(i)/V_{ip}(i)$ suggests that the system is closer to the 
error catastrophe point, where the stability condition in the distribution function 
$P(x_i=phenotype, a=genotype)$ is lost.
This leads to the increase in plasticity, with which the adaptation to a novel condition is achieved.
Once the fitted phenotypes are generated by this adaptation,
the variances decrease, with restoring  the proportionality between
$V_{g}(i)$ and $V_{ip}(i)$.


To sum up, adaptation to novel environment is characterized by the phenotypic variances as follows. 
With the increase in $V_{ip}(i)$,
the sensitivity of the phenotype to noise and  environmental change is increased,
thereby increasing plasticity.  
With the increase in $V_{g}(i)$, changeability of the phenotype by mutation is increased, thereby accelerating evolutionary change of each expression level.
With the increase in $V_g(i)/V_{ip}(i)$, the sensitivity to genetic change
is further increased, thus facilitating the evolution.  
With these trends---the increase in $V_{ip}(i)$, $V_g(i)$, $\rho_i=V_g(i)/V_{ip}(i)$, the adaptation to a novel environment is fostered.  
With this increase in plasticity,
gene expression dynamics
for adapting to a novel condition are explored. Once these fitted dynamics are shaped, 
the variances decrease, leading to a decrease in plasticity and increase in robustness.  


\subsection{Optimal noise level for varying environment}

\begin{figure}[tbp]
\begin{center}
\includegraphics[width=8cm,height=6cm]{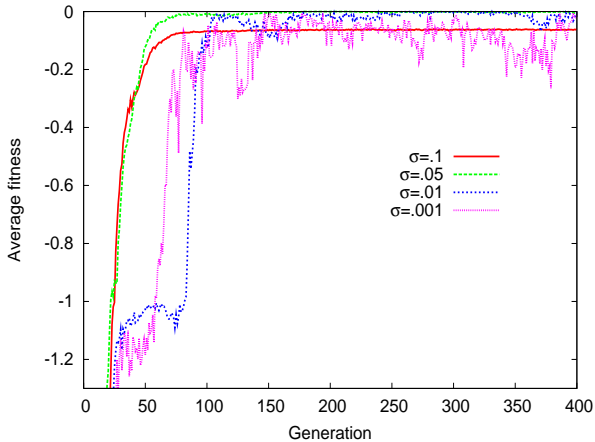}
\caption{The average of the mean fitness $<\overline{F}>$ plotted for each generation, under continual environmental variation, as described in the text. $M=64$, $k=8$, and $k_{inp}=4$,
while $I_j$ are changed
randomly within [0,.8].  The average of the mean fitness, $\overline{F}$, of each individual (over $L = 100$ runs) is computed over the total population
($N = 100$) at each generation.
The noise level, $\sigma$, is 0.1 (red),  0.05 ($\sim \sigma_c$; green), 0.01(blue), and 0.001 (pink).  At around $\sigma \sim 0.05$, the average fitness reaches the highest level.
}
\end{center}
\end{figure}

\begin{figure}[tbp]
\begin{center}
\includegraphics[width=8cm,height=6cm]{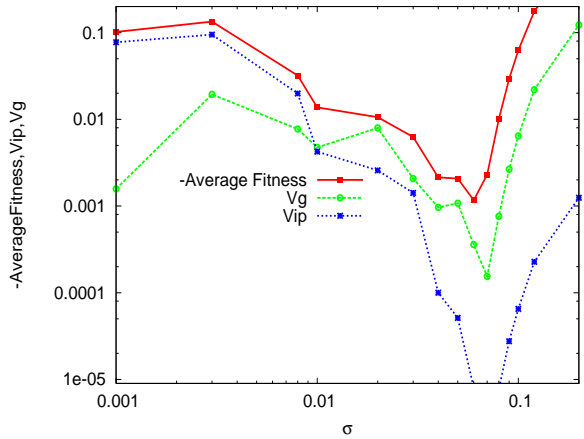}
\caption{The average fitness and the variances, $V_{ip}$ and $V_g$, through the course of the evolution under environmental variation as in Fig.6.  Each value is further averaged over 200-700 generations.
The overall temporal averages are plotted against the noise level $\sigma$.
Instead of the fitness itself, its sign inversion $-<\overline{F}>$ is plotted.
 At around $\sigma \sim 0.05$, the average fitness takes a maximal value, and
at $\sigma$ slightly below it, $V_g$ approaches $V_{ip}$.
}
\end{center}
\end{figure}

\begin{figure}[tbp]
\begin{center}
\includegraphics[width=8cm,height=6cm]{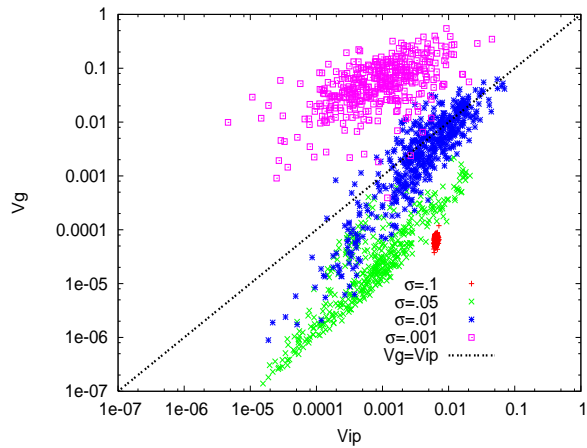}
\caption{The variances of the fitness ($V_{ip}$, $V_g$) through the course of the evolution under environmental variation as in Fig.6.
Each point is a result of one generation, and the plot is
taken over 200--700 generations.
The noise level $\sigma$ is 0.1 (red), 0.005 ($\sim \sigma_c$; green), 0.01 (blue), and 0.001(pink).
}
\end{center}
\end{figure}

Now, we consider evolution under continual environmental variation.
To discuss a long-term environmental change, we switch the environmental condition by generation. To be specific,  
we change randomly $I_i$ within $[0,1]$ per generation.

When environmental changes are continuously repeated, the decrease and increase  in
the variances $V_{ip}$ and $V_g$ are repeated.
Note that it takes more generations to adapt to a new fitness condition, if the phenotypic
variances have been smaller.
In our model, if the noise level in development is larger,
the phenotypic variances already take a small value
during the adaptation to satisfy the fitness condition.  Hence, in this case,
it takes more generations to adapt to a new fitness condition.  On the other hand, if the noise level $\sigma$ is smaller
than $\sigma_c \sim .05$, robust evolution does not progress.  Hence, for continuous
environmental change, there will be an optimal noise level to both adapt sufficiently fast to a new environment and
evolve the robustness of fitness for each environmental condition.  In Fig.6, we have plotted the
time course of the average fitness in population.
If the noise level is large, the system cannot follow the frequent environmental change and the
average fitness cannot increase sufficiently.
On the other hand, if the noise level is small, the fitness increases; however, if it is too small, the fitness of some individuals remains rather low.
Indeed, there is an optimal noise level at which the average fitness is maximal, as shown in Fig.7,
where the average fitness over generations is plotted against the noise level $\sigma$. This optimal noise level is close to the value of the
robustness transition   $\sigma_c$.

Next, we plotted the variances $V_{ip}$ and $V_g$ over generations (see Fig.8).  When $\sigma<\sigma_c$, then $V_g>V_{ip}$ and both the variances remain rather large, demonstrating that
robustness has not evolved at all.
For $\sigma \gg \sigma_c$, $V_g <V_{ip}$ and the variances remain small.  The robustness has evolved, but the system cannot adapt to an environmental change as the variances have become too small.
In contrast, for $\sigma\sim \sigma_c$, $V_{ip}$ and $V_g$ vary between low and high values over generations, maintaining the proportionality between the two variances, with $V_{ip}$ slightly larger than $V_{g}$.

\subsection{Adaptation against switches of the fitness condition}

\begin{figure}[h]
\begin{center}
\includegraphics[width=8cm]{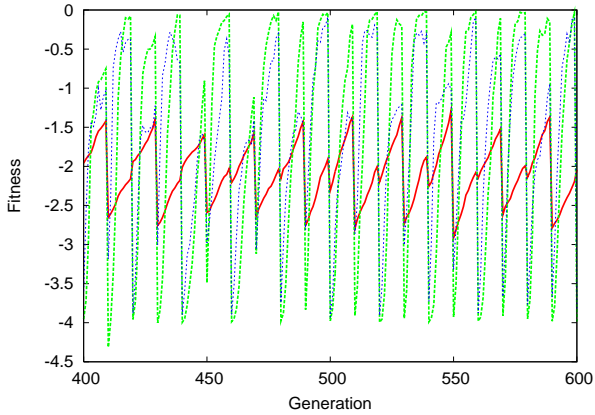}
\end{center}
\caption{The average of fitness plotted per generation,
where the fitness condition in the model (1)-(2) is
switched for every 10 generations between  $++++++++$ and $++++----$.  The average of the mean fitness $\overline{F}$ of each individual (over $L=200$ runs) is computed over the total population ($N=200$) at each generation.
The noise level $\sigma$ is 0.1 (red), 0.008 ($\sim \sigma_c$; green)  and 0.001 (blue).
}
\end{figure}

\begin{figure}[h]
\begin{center}
\includegraphics[width=8cm]{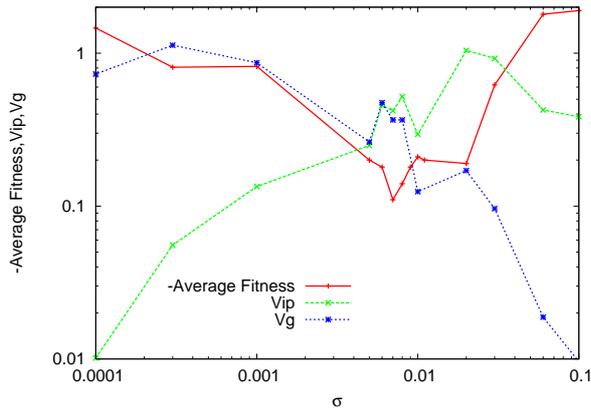}
\end{center}
\caption{The temporal average of the average fitness and the variances
$V_{ip}$ and $V_g$ in the evolution, with the change in the fitness condition for every 10 generations
as in Fig.9.
The overall temporal averages over population and over generations are plotted against the noise level $\sigma$.
Instead of the fitness itself, its sign inversion $-<\overline{F}>$ is plotted.
The temporal average is taken over 500--1000 generations.  At around $\sigma \sim .008$, the average fitness takes a maximal value, and at $\sigma$
slightly below it, $V_g$ exceeds $V_{ip}$.
}
\end{figure}

\begin{figure}[h]
\begin{center}
\includegraphics[width=8cm]{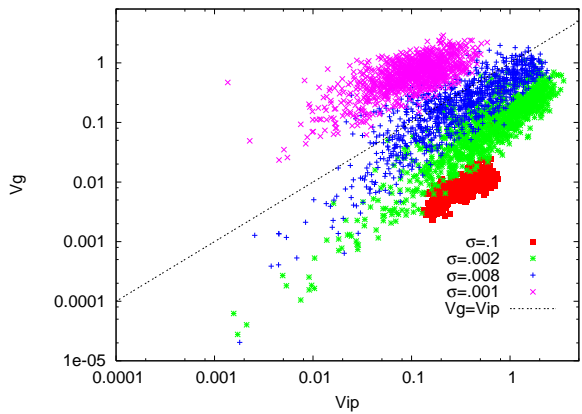}
\end{center}
\caption{Variances of the fitness, ($V_{ip}$, $V_g$), are plotted over generations through the course of the evolution with fitness change after every 10 generations,
as described in Fig.9.  Each point is a result of one generation, and the plot is
taken over 500--1000 generations.
The noise level $\sigma$ is 0.1 (red), 0.02(green), .008 ($\sim \sigma_c$;blue),  and 0.001 (pink).}
\end{figure}

For confirmation of the result in the last section,
we also carried out numerical experiments 
by adopting a separate procedure, i.e., by switching the
fitness condition.
As a specific example, we carried out the simulation by taking the model (1)-(2), without
including input terms (4).  After the gene expression
dynamics are evolved with the fitness to prefer $x_i>0$ for the target genes $i=1,2,...k(=8)$ adopted already,
then at a certain generation, we change the fitness condition
so that the genes $i=1,2,..,k/2$ are on and the rest are off (i.e., the fittest gene expression pattern is
$++++----$, instead of $++++++++$: In this model, the gene is off if $x_i <0$).  Here, we switch after sufficiently large generations when the fittest networks are evolved
(i.e., with $x_i>0$ for target genes).

In this case as well,  the variances of the fitness,
 $V_{ip}$ and $V_g$, first increase in proportion to adapt to a new fitness condition.
Later, they decrease in proportion to gain robustness to noise and mutation.
Next, we again computed $V_{ip}(i)$ and $V_g(i)$ successively through the course of the evolution.
Immediately after the switch in the fitness, the variances  $V_{ip}(i)$ and $V_g(i)$
increase as well as the ratio $V_g(i)/V_{ip}(i)$.  These variances in gene expression levels facilitate plasticity, and adaptation to a new environment.

When environmental changes are continuously repeated, the decrease and increase processes of
the variances $V_{ip}$ and $V_g$ are repeated.  In Fig.9, we have plotted the
time course of the average fitness in population, when the fitness condition is switched after every 10 generations.  In this case again, when the noise level is near $\sigma_c$, fast adaptation to a new environment and the increase of robustness at later generations are compatible.
Dependence of the average fitness on the noise level is shown in Fig.10, which also shows
an optimal noise level near the robustness transition   $\sigma_c$ (which is lower than the case in \S 5.2).  Indeed, below this noise level, $V_g$ exceeds $V_{ip}$ and the robustness to mutation is lost.
The plot of the variances $V_{ip}$ and $V_g$ over generations in  Fig.11 shows
that at $\sigma \sim \sigma_c$, they go up and down, maintaining an approximate proportionality between the two, with $V_{ip}$ slightly less than $V_{g}$.  Overall, the behavior here under the fitness switch agrees well with that under the environmental variation in \S 5.2.

\section{Summary and Discussion}

In the present paper, we have studied biological robustness and plasticity, in terms of phenotypic fluctuations.   The results are summarized into three points.

(1) Confirmation of earlier results on the phenotypic variances due to noise and due to genetic variation: The two variances, $V_{ip}$ (due to noise) and $V_g$ (due to mutation) decrease in proportion through the course of robust evolution under a fixed environmental condition.  After the evolution to achieve robustness to noise and mutation is completed, $V_{ip}(i)$ and $V_g(i)$ are proportional across expressions of most genes (or different phenotypic traits).  In short,

{\bf plasticity (changeability) of phenotype $\propto$ $V_{ip}$ $\propto$ $V_g \propto$ evolution speed}

\noindent
through the course of the evolution and across phenotypic traits (expressions of genes).

(2) Increase in the phenotypic variances and recovery of plasticity:  When robustness is increased under a given environmental condition, the system loses plasticity to adapt to a novel environment.  When the environmental condition is switched, both the phenotypic variances due to noise and due to genetic variation increase to gain plasticity and thus to adapt to the novel environment.  This increase is observed both for the variance of fitness and of each gene expression level.

(3) Optimal noise level to achieve both robustness and plasticity under continuous environmental change. There is generally a threshold level of noise in gene expression (or in developmental dynamics), beyond which robustness to noise and mutation evolves.  If the noise level is larger,
however, the system loses plasticity to adapt to environmental changes, whereas if it is much lower,
a robust, fitted phenotype is not generated.  At around the noise level for the "robustness transition," the system can adapt
to environmental changes and achieve a higher fitness.
There, the phenotypic variances $V_{ip}$ and $V_g$ increase and decrease, roughly
maintaining the proportionality between the two, while sustaining $V_{ip} \stackrel{>}{\sim} V_g$.

From a statistical-physicist viewpoint, the relationship between responsiveness and fluctuation is expected as a proper extension of the fluctuation-response relationship.  In this context, it is rather natural that the response ratio to environmental change, i.e., plasticity, is proportional to $V_{ip}$,
the isogenic phenotypic fluctuation.
On the other hand, as Fisher stated, the variance due to genetic change, $V_g$, is proportional to evolution speed, i.e., response of the fitness against mutation and selection.  Interestingly, our simulations and evolutionary stability argument suggest the proportionality
between $V_{ip}$ and $V_g$.  This implies the proportionality between environmental plasticity and evolvability.

In fact, Waddington\cite{Waddington,Schmalhausen} coined the term genetic assimilation, in which phenotypic changes induced by environmental changes foster later genetic evolution. Since then, positive roles of phenotypic plasticity in evolution have
been extensively discussed\cite{West-Eberhard,Kirschner,AncelFontana}.
Our study gives a quantitative representation of such relationship in terms of fluctuations. 

Existence of the threshold noise level below which
 robustness is lost is reminiscent of a glass transition in physics:
 For a higher noise level, dynamical systems
for global attraction to a functional phenotype are generated through evolution,
whereas for a lower noise level,  the dynamics follow motion in a rugged landscape, where perturbation to it leads to a failure in the shaping of the functional phenotype. In fact, Sakata et al. considered
a spin-glass model whose interaction matrix evolves to generate a high-fitness thermodynamic state.  The transition to lose robustness was found by lowering the temperature. Interestingly,
this transition is identified as the replica symmetry breaking transition from a replica symmetric phase\cite{Sakata,Sakata2}.

Under environmental fluctuation, the evolution to achieve both plasticity to a new environment and robustness of a fitted state is possible near this transition, for losing the robustness.  
In other words, one may regard that a biological systems favors the "edge-of-glass" state.

\section*{Acknowledgements}
 I would like to thank T. Yomo, C. Furusawa, M. Tachikawa, A. Sakata, K. Hukushima, L. Lafuerza and S. Ishihara
 for stimulating discussion.
This work was supported by a Grant-in-Aid for Scientific Research
(No 21120004) on Innovative Areas ``The study on the neural dynamics for understanding
communication in terms of complex hetero systems (No.4103)'' of MEXT, Japan.


\begin{thebibliography}{999}

\bibitem{Evolution}
de Visser JA, et al. (2003)
Evolution and detection of genetic robustness.
{\sl Evolution} {\bf 57}:1959-1972.

\bibitem{Callahan}
Callahan HS, Pigliucci M., and Schlichting CD
(1997) Developmental phenotypic plasticity: where ecology and evolution meet molecular biology.
{\sl Bioessays}{\bf 19}:519-525.

\bibitem{West-Eberhard}
West-Eberhard MJ (2003)
{\sl Developmental Plasticity and Evolution} (Oxford University Press).

\bibitem{Kirschner}
Kirschner MW, and Gerhart JC (2005),
{\sl The Plausibility of Life} (Yale University Press).

\bibitem{AncelFontana}
Ancel LW, and Fontana W (2002)
Plasticity, evolvability, and modularity in RNA.
{\sl J. Exp. Zool.} {\bf 288}:242-283.

\bibitem{Frank}
Frank, S.A., (2012)
{\sl Natural selection. II. Developmental variability and evolutionary rate*},
{\sl Journal of Evolutionary Biology},
{\bf 24}, 2310--2320,

\bibitem{Wagner}
Wagner A (2002)
Robustness against mutations in genetic networks of yeast.
{\sl Nature Genetics} {\bf 24}:355-361.

\bibitem{Wagner2}
Wagner A. 2005, {\sl Robustness and evolvability in living systems},
Princeton University Press, Princeton NJ.

\bibitem{Wagneretal}
Wagner GP, Booth G, Bagheri-Chaichian H, 1997,
A population genetic theory of canalization.
{\sl Evolution} {\bf 51}:329-347.

\bibitem{Barkai}
Barkai N and Leibler S (1997)
Robustness in simple biochemical networks.
{\sl Nature}, {\bf 387}:913-917.

\bibitem{Alon}
Alon U, Surette MG, Barkai N, Leibler S
(1999) Robustness in bacterial chemotaxis.
{\sl Nature}, 1999, {\bf 397}:168-171.

\bibitem{Bergman}
Siegal ML and Bergman A (2002)
Waddington's canalization revisited: Developmental stability and evolution.
{\sl Proc Nat. Acad. Sci. U S A} {\bf 99}:10528-10532.

\bibitem{Ciliberti}
Ciliberti S, Martin OC, Wagner A (2007)
Robustness can evolve gradually in complex regulatory gene networks with varying topology.
{\sl PLoS Comp. Biology} {\bf 3}:e15.

\bibitem{Plos1}
Kaneko K (2007)
Evolution of robustness to noise and mutation in gene expression dynamics.
{\sl PLoS ONE} {\bf 2}:e434.


\bibitem{JTB}
Kaneko K and Furusawa C (2006)
An evolutionary relationship between genetic variation and phenotypic fluctuation.
{\sl J. Theo. Biol.} {\bf 240}:78-86.

\bibitem{book}
Kaneko K (2006) {\sl Life: An Introduction to Complex Systems Biology} (Springer, Heidelberg and New York).

\bibitem{Sato}
Sato K, Ito Y, Yomo T, Kaneko K (2003)
On the relation between fluctuation and response in biological systems.
{\sl Proc. Nat. Acad. Sci. U S A} {\bf 100}:14086-14090.

\bibitem{LehnerKK}
Lehner B. and Kaneko K., 2011,
A macroscopic relationship between fluctuation and response in biology.
{\sl Cellular and Molecular Life Sciences}, {\bf 68}, 1005-1010

\bibitem{Fisher}
Fisher RA (1930) {\sl The Genetical Theory of Natural Selection}, (Oxford University Press).

\bibitem{Futuyma}
Futuyma DJ (1986) {\sl Evolutionary Biology} (Second edition), (Sinauer Associates Inc., Sunderland).

\bibitem{Hartl}
Hartl DL and Clark AG (2007)
{\sl Principles of Population Genetics}
(Sinauer Assoc. Inc., Sunderland) 4th ed.

\bibitem{Price}
Price, G. R. (1970). Selection and covariance. Nature
{\bf 227}:520-521.

\bibitem{Lande}
 Lande R., and  Arnold S.J., (1983)
{\sl The Measurement of Selection on Correlated Characters}
Evolution 37:  1210-1226

\bibitem{Ao}
Ao P., (2005)
{\sl Laws in Darwinian evolutionary theory},
Physics of Life Reviews,
{\bf 2}: 117-156

\bibitem{Mustonen}
Mustonen V., and Lassig M.,(2010)
Fitness flux and ubiquity of adaptive evolution,
{\sl Proc. Nat. Acad. Sci. U S A}, 107, 4248-4253 

\bibitem{Elowitz}
Elowitz MB, Levine AJ, Siggia ED, Swain PS (2002)
Stochastic gene expression in a single cell.
{\sl Science} {\bf 297}:1183-1187.

\bibitem{Bar-Even}
Bar-Even A, et al. (2006)
Noise in protein expression scales with natural protein abundance.
{\sl Nature Genetics} {\bf 38}:636-643.

\bibitem{Kaern}
Kaern M, Elston TC, Blake WJ, Collins JJ (2005) Stochasticity in gene expression: From theories to phenotypes.
{\sl Nat. Rev. Genet.} {\bf 6}:451-464.

\bibitem{Furusawa}
Furusawa C, Suzuki T, Kashiwagi A, Yomo T, Kaneko K (2005)
Ubiquity of log-normal distributions in intra-cellular reaction dynamics.
{\sl Biophysics} {\bf 1}:25-31.

\bibitem{Tsuru}
Tsuru S., Ichinose J., Sakurai T.,Kashiwagi A., Ying B-W., Kaneko K., and Yomo T.,
"Noisy cell growth rate leads to fluctuating protein concentration in bacteria"
Physical Biology 6 (2009) 036015

\bibitem{Wakamoto}
Wakamoto Y. Ramsden J. and Yasuda K., (2005)
Single-cell growth and division dynamics showing epigenetic correlations 
Analyst, {\bf 130},
 311-317

\bibitem{Wright}
Wright, S., (1932)
{\sl The roles of mutation, inbreeding, crossbreeding and selection in evolution},
{\sl Proceedings of the sixth international congress on genetics},
{\bf 1}, 356--366,


\bibitem{Sakata}Sakata A, Hukushima K, Kaneko K (2009)
Funnel landscape and mutational robustness as a result of evolution under thermal noise.
{\sl Phys. Rev. Lett.} \textbf{102}: 148101.

\bibitem{Sakata2} Sakata A, Hukushima K, Kaneko K (2011), Replica symmetry breaking in an adiabatic spin-glass model of adaptive evolution,
arXiv1111.5770v1, submitted to Europhys. Lett.

\bibitem{Kaneko2009}
Kaneko K (2009)
Relationship among phenotypic plasticity, genetic and epigenetic fluctuations, robustness, and evolvability.
\emph{J. BioSci}. \textbf{34}:529-542.


\bibitem{gene-net}
Glass L and Kauffman SA (1973)
The logical analysis of continuous, non-linear biochemical control networks.
{\sl J. Theor. Biol.} {\bf 39}:103-129.

\bibitem{Mjolsness}
Mjolsness E, Sharp DH, Reisnitz J (1991)
A connectionist model of development.
{\sl J. Theor. Biol.} {\bf 152}:429-453

\bibitem{Sole}
Salazar-Ciudad I, Garcia-Fernandez J, and Sole, RV (2000)
Gene networks capable of pattern formation: from induction to reaction--diffusion.
{\sl J. Theor. Biol.} {\bf 205}: 587-603.

\bibitem{Chaos}
Kaneko K. (2008)
Shaping robust system through evolution.
{\sl Chaos} {\bf 18}:026112.

\bibitem{BMC}
Kaneko K. (2011) Proportionality between variances in gene expression induced by noise and mutation: consequence of evolutionary robustness. {\sl BMC Evol Biol} {\bf 11}:27

\bibitem{Luis}
Lafuerza, L.F.  and Kaneko K., in preparation.

\bibitem{Onuchic}
Onuchic J.N., Wolynes P.G., Luthey-Schulten Z., and Socci
N.D. (1995) ``Toward an Outline of the Topography of a Realistic
Protein-Folding Funnel", \emph{Proc. Nat. Acad. Sci. USA} {\bf 92}, 3626.


\bibitem{Li}
Li, F., Long, T., Lu, Y., Ouyang, Q., and Tang, C. (2004)
"The yeast cell-cycle network is robustly designed',
{\sl Proc. Nat. Acad. Sci. U S A}, {\bf 101}, pp. 10040--10046

\bibitem{Abe}
Abe H. and Go. N. (1980) " Noninteracting local-structure model of
folding and unfolding transition in globular proteins. I.
Formulation", \emph{Biopolymers} {\bf 20}, 1013.

\bibitem{Eigen}
Eigen M. and  Schuster P. (1979) {\sl The Hypercycle}, (Springer, Heidelberg).

\bibitem{Laundry}
Landry CR, Lemos B, Rifkin, SA, Dickinson WJ, and Hartl DL (2007)
Genetic Properties Influencing the Evolvability of Gene Expression.
\emph{Science} \textbf{317}: 118.

\bibitem{LehnerPLoS}
Lehner B  (2010) Genes Confer Similar Robustness to Environmental, Stochastic, and Genetic Perturbations in Yeast.
{\sl PLoS One} e9035.

\bibitem{Stearns}
Stearns SC, Kaiser M, Kawecki TJ (1995)
The differential genetic and environmental canalization of fitness components in Drosophila melanogaster.
\emph{J. Evol. Biol.} \textbf{8}:539-557.

\bibitem{Waddington}
Waddington CH (1957)
{\sl The Strategy of the Genes} (Allen \& Unwin, London).

\bibitem{Schmalhausen}
Schmalhausen II  (1949, reprinted 1986)
{\sl Factors of Evolution: The Theory of Stabilizing Selection}
(University of Chicago Press, Chicago).

\bibitem{envfluct1}
Slatkin, M. and Lande, R.
1976, {\sl Niche width in a fluctuating environment-density independent model},
American Naturalist, 31-55,

\bibitem{Kussel}
Leibler S. and Kussel E., (2010)
{\sl Individual histories and selection in
heterogeneous populations}
{\sl Proc. Nat. Acad. Sci. U S A} 107,  13183-13188

\bibitem{envfluct2}
Rivoire, O. and Leibler, S., (2011)
{\sl The value of information for populations in varying environments},
Journal of Statistical Physics,
{\bf 142},
1124--1166

\bibitem{Mustonen2}
Mustonen V., and Lassig M.,(2007)
Adaptations to fluctuating selection in {\sl Drosophila}, 
{\sl Proc. Nat. Acad. Sci. U S A}, 104, 2277-2282


\end{thebibliography}
\end{document}